\documentclass[aps,superscriptaddress,twocolumn,amssymb,longbibliography]{revtex4-1}
\usepackage{amsmath}
\usepackage{amssymb}
\usepackage{graphicx}
\usepackage{braket}
\usepackage{hyperref}
\usepackage{hyphenat}
\begin{document}
\title{Realization of a binary-outcome projection measurement of a three-level superconducting quantum system}

\author{Markus Jerger}
\affiliation{ARC Centre of Excellence for Engineered Quantum Systems, The University of Queensland, St Lucia QLD 4072
Australia}

\author{Pascal Macha}
\affiliation{ARC Centre of Excellence for Engineered Quantum Systems, The University of Queensland, St Lucia QLD 4072
	Australia}

\author{Andr\'es Rosario Hamann}
\affiliation{ARC Centre of Excellence for Engineered Quantum Systems, The University of Queensland, St Lucia QLD 4072
	Australia}

\author{Yarema Reshitnyk}
\affiliation{School of Mathematics and Physics, University of Queensland, Brisbane, Queensland 4072, Australia}

\author{Kristinn Juliusson}
\affiliation{Quantronics, SPEC, IRAMIS, DSM, CEA Saclay, Gif-sur-Yvette, France}

\author{Arkady Fedorov}
\email{a.fedorov@uq.edu.au}
\affiliation{ARC Centre of Excellence for Engineered Quantum Systems, The University of Queensland, St Lucia QLD 4072
	Australia}
\affiliation{School of Mathematics and Physics, University of Queensland, Brisbane, Queensland 4072, Australia}

\begin{abstract}

Binary-outcome measurements allow to determine whether a multi-level quantum system is in a certain state while preserving quantum coherence between all orthogonal states. In this paper, we explore different regimes of the dispersive readout of a three-level superconducting quantum system coupled to a microwave cavity in order to implement binary-outcome measurements. By designing identical cavity frequency shifts for the first and second excited states of the system, we realize strong projective binary-outcome measurements onto its ground state with a fidelity of $94.3\%$. Complemented with standard microwave control and low-noise parametric amplification, this scheme enables the quantum non-demolition detection of leakage errors and can be used to create sets of compatible measurements to reveal the contextual nature of superconducting circuits.
\end{abstract}


\date{\today}

\maketitle


\section*{Introduction}
Measurements that discriminate in which subspace the state vector of a system lies while preserving coherence within all subspaces play a pervasive role in quantum mechanics. Such degenerate measurements are required to demonstrate Kochen-Specker contextuality~\cite{Szangolies2015} and are used extensively in quantum error correction~\cite{Terhal2015}, magic state distillation~\cite{Bravyi2005}, and entanglement distillation~\cite{Bennett1996}. However, in applications related to error correction, the degenerate measurements of the original system are typically realized by fully projective measurements of a coupled ancilla, which are more straightforward to implement. In this paper we realize a direct degenerate measurement of an indivisible superconducting qutrit using generic properties of dispersive readout~\cite{Blais2004}. By engineering the dispersive shifts of a cavity coupled to the qutrit, our measurement apparatus is tuned to be highly selective on one of the basis states but fully insensitive to the other two basis states of the qutrit. With tomographical reconstruction of the states before and after measurement, we experimentally prove the protection of the coherence between the first and second excited states of the qutrit during the measurement, thus demonstrating the most fundamental property of degenerate measurements. As an immediate application, our experiment enables the quantum non-demolition measurement of 'leakage errors'~\cite{Suchara2015} and Kochen-Specker tests of quantum contextuality~\cite{Klyachko2008, Yu2012} with indivisible superconducting qutrits.

A binary-outcome projective measurement can be formulated in the form of a question: {\it Is the system in state $|v\rangle$ or not}? Such a measurement is described by a set of two measurement operators $\{M_{1,2}\}=\{|v\rangle\langle v|, I - |v\rangle\langle v|\}$. Assigning measurement outcomes $\{m_{1,2}\}$ to $\pm1$, the measurement can be associated~\cite{Nielsen2000} with the binary observable $A_v = \sum_{j} m_i M_i = 2 |v\rangle\langle v| - I$. The measurement observable alone is sufficient to evaluate the outcome statistics. However, while an observable is fully determined by the set measurement operators, the reverse is not correct and different sets of measurement operators can be associated with the same observable. To evaluate the effect of the measurement on the system it is necessary to know the specific measurement operators as the density operators before and after the measurement are connected by $\rho_{\rm a} = \sum_{i}M_{i}\rho_{\rm b} M^\dagger_{i}$.

For systems with more than two dimensions the binary-outcome measurement is an example of a degenerate measurement: if the system is {\it not} found in state $|v\rangle$, its coherence is fully preserved. This distinguishing feature of the degenerate measurement is crucial to a number of applications. In particular, if two states (or rays) $\{|v\rangle,|u\rangle\}$ are orthogonal ($\langle v|u\rangle = 0$), the measurement operators and observables associated with the corresponding binary-outcome measurement commute. Such measurements are called {\it compatible} -- they can be measured simultaneously (or in any order) without disturbing each other. Compatibility is central for demonstrating the contextual nature of quantum mechanics~\cite{Szangolies2015}, which is intimately related to the power of quantum computing~\cite{Howard2014}.

To be specific, let us consider a three-level system spanned by the orthonormal logical eigenbasis $\{|0\rangle, |1\rangle, |2\rangle\}$ (for brevity assume it is also the energy eigenbasis). The binary-outcome measurement testing whether the qutrit is in its ground state is given by measurement operators $\{M_{+1},M_{-1}\}=\{|0\rangle\langle 0|,I - |0\rangle\langle 0|\} $ associated with outcomes $\{+1,-1\}$, respectively, and is described by the corresponding binary observable $A_{0}$. The density operator after the measurement reads $\rho_{\rm a} =\sum_{i=\pm1}M_{i}\rho_{\rm b} M^\dagger_{i}= \rho_{\rm b, 00}|0\rangle\langle 0| +\sum_{i,j=1}^{2}\rho_{\rm b, ij}|i\rangle\langle j|$, proving that any quantum coherence associated with $|0\rangle$ is fully lost while coherence in the orthogonal subspace spanned by $\{|1\rangle, |2\rangle\}$ is preserved. By rotating the quantum state before and after the measurement using standard quantum control techniques, one can implement projection on any state.

 
In practice, the readout for a physical qutrit may realize a ternary measurement when all three states $\{|0\rangle,|1\rangle, |2\rangle\}$ are resolved and the measurement is described by the three operators $\{|0\rangle\langle 0|, |1\rangle\langle 1|, |2\rangle\langle 2|\}$. By nominally assigning the outcomes for projectors $|1\rangle\langle 1|$ and $|2\rangle\langle2|$ to $-1$ and  $|0\rangle\langle0|$ to $+1$, the measurement can be associated with the same binary observable $A_{0} = \sum_{i=1,2,3} m_i M_i=2 |0\rangle\langle 0|-I$, and the outcomes will be identical to the binary-outcome measurement described by $\{|0\rangle\langle 0|,I - |0\rangle\langle 0|\}$.  However, for the ternary measurement the coherence between $|1\rangle$ and $|2\rangle$ is lost, making this readout scheme unsuitable of a number of quantum protocols.

In this paper we use a multi-level superconducting quantum system coupled to a microwave cavity to engineer a measurement apparatus which is highly selective on the ground state, but is fully insensitive to the other excited states. As no information about the relative populations of the excited states is extracted, their mutual coherence is not affected by measurement back-action. This property allows us to realize a near perfect binary-outcome measurement for a superconducting qutrit.

Our manuscript is organized as follows. In Section 2 we present the theory of a multilevel quantum system coupled to a microwave cavity in the dispersive regime. We introduce the state-dependent dispersive shifts of the cavity frequency and explain how they can be used to realize a quantum measurement. As the dispersive shifts for the first and second excited states are in general different, we speculate that the standard readout will result in a full collapse of quantum coherence. We then outline the conditions for these shifts to be identical and the implications for the readout. In Section 3 we present our main experimental results. First, we give the details of our experimental setup. Second, we implement a procedure to identify the conditions where the readout does not affect the coherence between the first and second excited states. Finally, we use a full tomographic reconstruction to extensively characterize the effect of the readout on the qutrit. In the last section we summarize our results and discuss the implications of our readout scheme for testing quantum contextuality and for error correction.

\section*{Dispersive readout of superconducting qutrits}
A standard superconducting quantum system of the transmon type has a weakly anharmonic multi-level structure whose three lowest energy eigenstates, labeled $\{|0\rangle,|1\rangle, |2\rangle\}$, are ideally suited to encode the logical states of a qutrit. The Hamiltonian of a qutrit coupled to a microwave resonator takes the generalised Jaynes-Cummings form~\cite{Koch2007,Boissonneault2010} $H = H_0 + \hbar\sum_{i=0,1} g_{i,i+1} \left(a^\dagger |i\rangle\langle i+1| + a |i+1\rangle\langle i|\right)$, where $H_0 = \hbar\omega_r a^\dagger a + \hbar\sum_{i=0,1,2} \omega_i |i\rangle\langle i|$ is the free Hamiltonian, $\omega_r$, $a$ and $a^\dagger$ are the frequency, creation and annihilation operators for the quantized resonator mode, $\omega_i$ is the frequency of level $|i\rangle$, and $g_{i,i+1}$ is the coupling strength between the resonator mode $a$ and the $i\leftrightarrow i+1$ transition of the qutrit.

\begin{figure*}[htbp]
\begin{center}
\begin{tabular}{ccc}
	\includegraphics[scale=1]{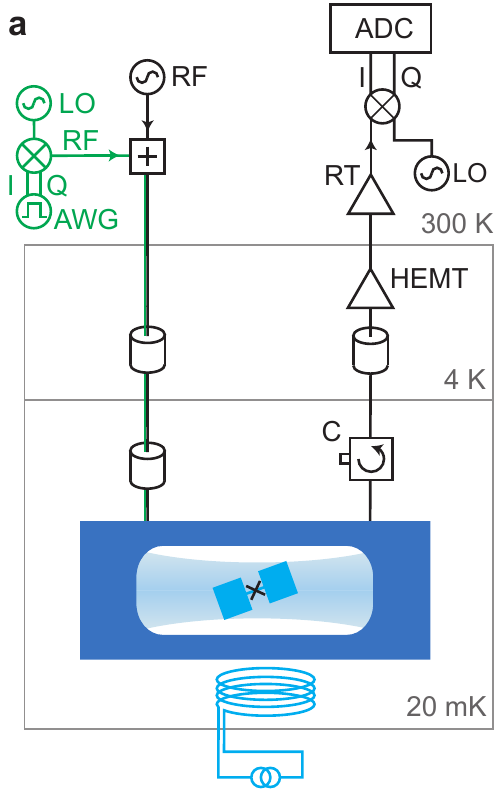} &
	\includegraphics[scale=1]{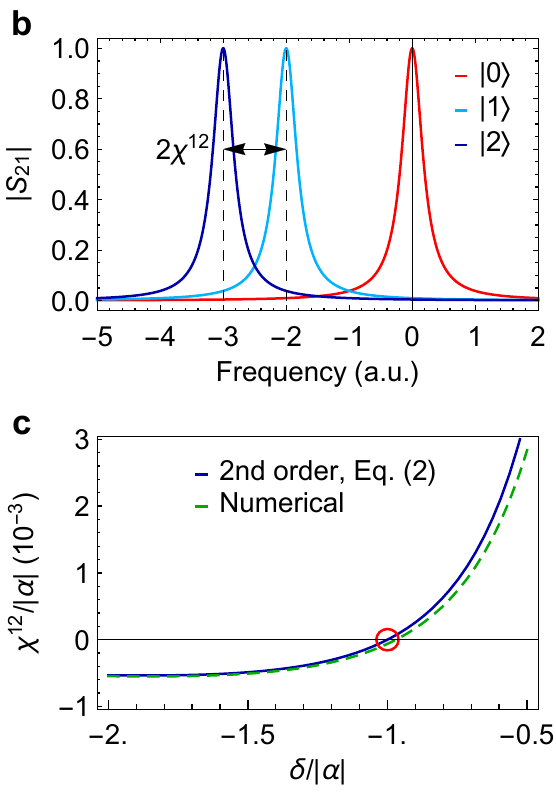} &
	\includegraphics[scale=1]{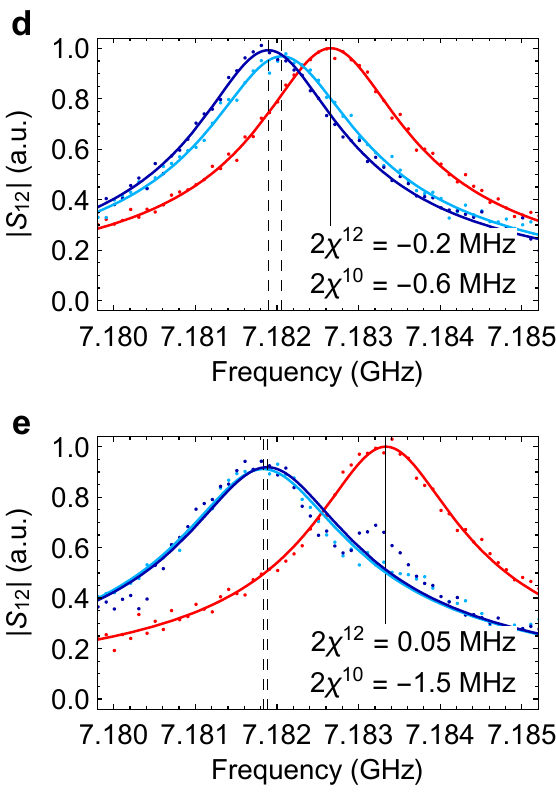}
\end{tabular}
\caption{
	(a) Simplified diagram of the measurement setup. A transmon type multi-level quantum system is incorporated into a 3D microwave copper cavity attached to the cold stage of a dilution cryostat. A magnetically tunable Josephson junction (SQUID) is used to control the transition frequency of the qutrit by a superconducting coil attached to the cavity. Amplitude-controlled and phase-controlled microwave pulses are applied to the input port of the cavity by a quadrature IF (IQ) mixer driven by a local oscillator (LO) and sideband modulated by an arbitrary waveform generator (AWG). The measurement signals transmitted through the cavity are amplified by a high-electron-mobility transistor (HEMT) amplifier at 4~K and a chain of room temperature (RT) amplifiers. The sample at 20~mK is isolated from the higher temperature stages by three circulators (C) in series. The amplified transmission signal is down-converted to an intermediate frequency of 25~MHz in an IQ mixer driven by a dedicated LO, and is digitized by an analogue-to-digital converter (ADC) for data analysis.
	(b) Sketch of the transmission through a cavity for a qutrit prepared in the ground (red), first excited (light blue) and second excited (dark blue) states. $\chi^{12}$ indicates the difference in dispersive shifts for the first and second excited states. 
	(c) Dispersive shift $\chi^{12}$ as a function of detuning based on the second order approximation (\ref{chi12}) (solid line) and on numerical diagonalization of the generalized Jaynes-Cummings Hamiltonian (dashed line). The red circle indicates the sweet spot where the readout cannot distinguish between the excited states. This condition can be used to realize a perfect binary-outcome projection on the ground state of a qutrit.
	(d) Pulsed spectroscopy of the microwave readout cavity for the qutrit prepared in different basis states off the sweet spot: transmission through the cavity for a qutrit prepared in the ground (red), first excited (light blue) and second excited (dark blue) states for $f_{01} = 6.750$~GHz and $\alpha = -310$~MHz. Dots are measured data and the solid lines are fit to the Lorentzian function. Both dispersive shifts $\chi^{01}$  and $\chi^{12}$ are of the same order. 
	(e) Cavity pulsed spectroscopy at the sweet spot, $f_{01} =  6.906$~GHz. The cavity lines for $|1\rangle$ and $|2\rangle$ are virtually indistinguishable. Note that the side-peak visible in the cavity response for the qutrit prepared in the excited states appears due to relaxation of the qutrit to the ground state during integration of the signal. 
} 
\label{fig1}
\end{center}
\end{figure*}

In the dispersive regime, $4 \langle a^\dagger a\rangle (g_{i,i+1}/\Delta_{i,i+1})^2\ll 1$, with $\Delta_{i+1,i} = \left(\omega_{i+1} - \omega_{i}\right) - \omega_r$, the Hamiltonian can be approximated~\cite{Boissonneault2010} by the diagonal Hamiltonian $H^D$ up to the second order in $g_{i,i+1}/\Delta_i$:
\begin{equation}\label{H^D}
H^D\approx \hbar \tilde\omega_r a^\dagger a + \hbar \sum_{i=0,1,2} (\tilde\omega_i + S_i a^\dagger a) |i\rangle\langle i|,
\end{equation}
where frequencies $\tilde\omega_i$ and $\tilde\omega_r$ include the Lamb shifts and the ac-Stark coefficients are defined as $S_i = \chi_{i-1,i} - \chi_{i,i+1}$, with $\chi_{i, i+1} = g_{i,i+1}^2/\Delta_{i,i+1}$ for $i\geq 0$ and $\chi_{-1, 0} = 0$~\cite{Koch2007}. The diagonal form of (\ref{H^D}) shows that the frequency of the cavity acquires a dependence on the state of the qutrit. 
The frequency shift between the ground and the first excited states $2\chi^{01} = S_1 - S_0 = 2\chi_{01} -\chi_{12}$ is extensively used for realizing dispersive readout of superconducting qubits by measuring microwave transmission through the cavity~\cite{Blais2004}. With recent advances in microwave amplification near the quantum limit~\cite{Castellanos2008}, dispersive readout became the most advanced and most common way to achieve single-shot quantum non-demolition readout of superconducting qubits~\cite{Riste2012}. 

Using Eq.~(\ref{H^D}) one can obtain the relative dispersive shift between the first and second excited states as
\begin{multline} \label{chi12}
2\chi^{12} = S_2-S_1=2\chi_{12} -\chi_{23} - \chi_{01} \\
=-\frac{2 g^2 \alpha (\alpha -\delta)}{\delta (\delta + \alpha) (\delta + 2 \alpha)},
\end{multline}
where we define the interaction strength and detuning relative to the primary transition of the system: $g = g_{01}$ and $\delta = \Delta_{01}$. Here we also used that $g_{i,i+1} = \sqrt{i+1}g$ and the anharmonicity $\alpha = (\omega_{i+1} -\omega_i) -(\omega_{i} -\omega_{i-1})$ is independent of $i$ as expected for $E_j/E_c \gg1 $, where $E_c$ is the charging energy of the transmon~\cite{Sete2015}.

$\chi^{12}$ is, in general, of the same order as the qubit dispersive shift $\chi^{01}$. It follows that the dispersive readout is capable of distinguishing all logical states of the qutrit~\cite{Bianchetti2010} (see Fig.~\ref{fig1}~b), leading to full dephasing of the system after the measurement. Yet the dependence on the detuning suggests that at the special point  $\delta = \alpha$ the dispersive shift  $\chi^{12}$ vanishes (Fig.~\ref{fig1}~c). This conclusion is confirmed by numerical diagonalization of the Hamiltonian taking all orders of $g/\delta$ into account. To do that we diagonalized the generalised Jaynes-Cummings Hamiltonian taking into account the six lowest states of the multi-level quantum system and six photon number states of the cavity.  Then we estimated the cavity frequency as the the difference between one and zero photon states for a given state of the qutrit. The results suggest that this "sweet spot" can be used to realize a perfect binary-outcome measurement of a qutrit. Similar in spirit, engineering of identical dispersive shifts for a two-qubit system was also used to create two qubits maximally entangled by measurement~\cite{Riste2013,Roch2014}.

\begin{figure*}[htbp]
\begin{center}
\includegraphics[scale=1]{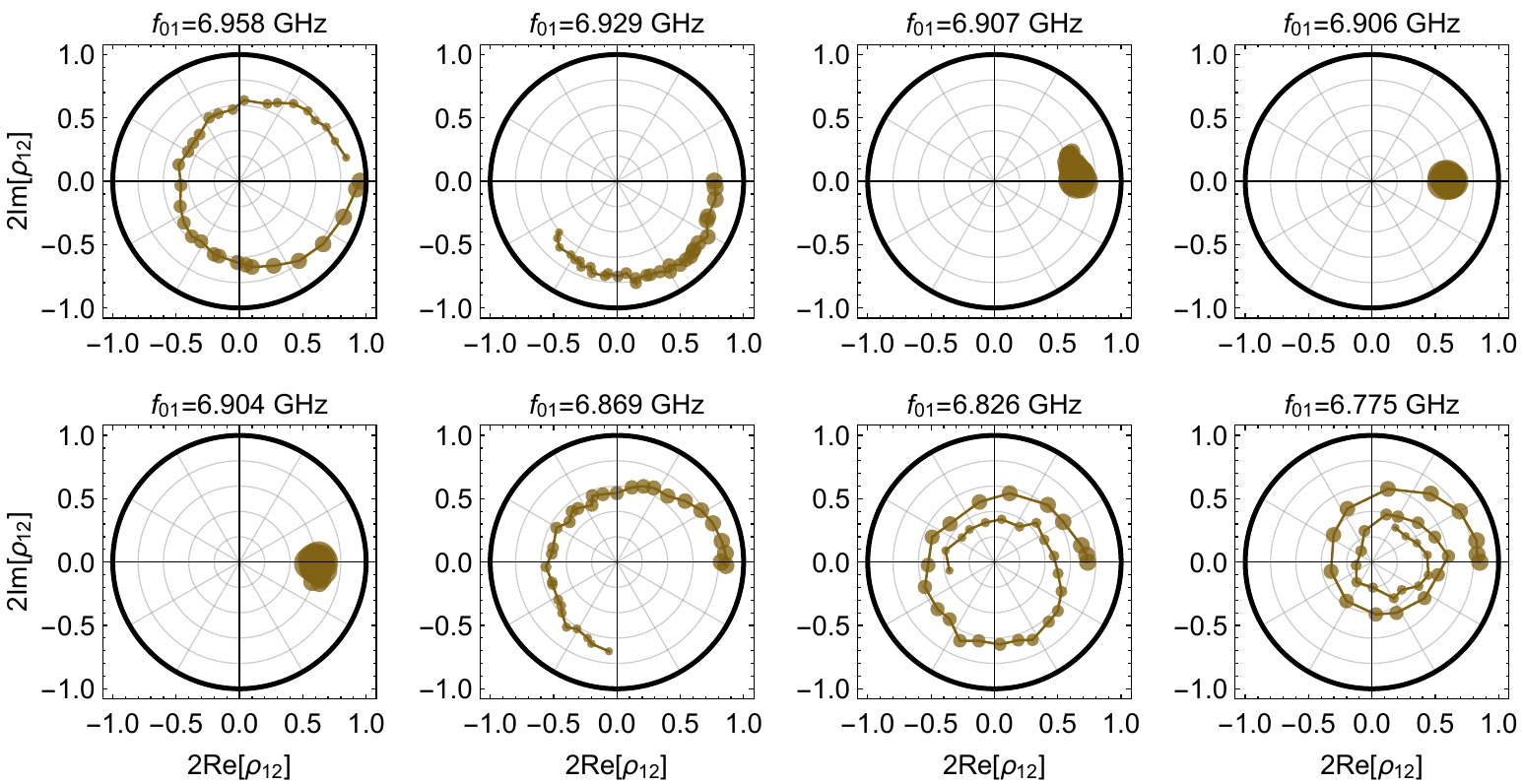}
\caption{
	Effect of the readout on the coherence of the $1\leftrightarrow2$ transition. The qutrit is prepared in $(1/\sqrt2)(|1\rangle + |2\rangle)$ and its coherences $\rho_{12}$ are tomographycally reconstructed after 2~$\mu$s. An extra readout pulse is inserted between the preparation and tomography pulses. For each frequency of the qutrit $f_{01}$ the length of the extra readout pulse was varied between 0 and 1.5~$\mu$s indicated as the radius of dots with larger dots corresponding to shorter pulse lengths. The readout pulse leads to an extra phase and to extra dephasing for the $1\leftrightarrow2$ transition manifested in a spiral trajectory towards the center of the plot. $f_{01}=6.906$~GHz was identified as the sweet spot, as there is no effect of the readout on $\rho_{12}$.} 
\label{find_sweet_spot}
\end{center}
\end{figure*}

\section*{Results}
\subsection{System}
To implement our binary-outcome readout scheme we use a transmon type multilevel quantum circuit incorporated into a 3D microwave copper cavity (Fig.~\ref{fig1}(a)). The circuit is fabricated on high resistivity Si ($>$ 1000~Ohm/cm) substrate  in a single step of electron beam lithography followed by shadow evaporation of two Al layers with thicknesses of 25~nm and 35~nm, respectively, with an oxidation step  between the depositions for 5~min at 7.0~mbar. The design of the circuit consists of two planar capacitor plates  with size of 700$\,\mu$m by $350\,\mu$m. The plates  separated by 50$\,\mu m$ and connected via a line interrupted by a micron size DC SQUID, playing the role of a magnetically tunable Josephson junction~\cite{Paik2011}. Magnetic flux supplied by a superconducting coil attached to the copper cavity is used to control the transition frequencies of the circuit. 

The  cavity is coupled asymmetrically to the input and output microwave ports with corresponding external quality factors of $Q_{in} \simeq 80\,000$ and $Q_{out} = 4\,200$ for transmission measurements. The internal quality factor of the cavity is expected to be $Q \sim 10\,000 - 15\,000$ at mK temperatures. The resonance frequency of the fundamental mode used for readout is $f_r\equiv \tilde{\omega}_r/(2 \pi) = 7.182$~GHz and the maximum primary transition of the qutrit $f_{01} \equiv(\tilde\omega_1 -\tilde\omega_0)/(2\pi) = 6.955$~GHz. The value of the anharmonicity of the qutrit, estimated as $\alpha\simeq -E_c\simeq -300$~MHz, allowed us to tune the qutrit to the sweet spot at $\delta\simeq\alpha$.
 
In most experiments the dipole moment of a superconducting quantum system is oriented in parallel to the electric field of the cavity mode to maximize system-resonator coupling. To enter the dispersive regime for our relatively small detuning we rotate the chip to 76 degrees with respect to the electric field orientation. At this angle we reach a coupling strength of $g =  20.0$~MHz and for $\delta\simeq -300$~MHz the cavity frequency pull is of the order of the cavity linewidth $2|\chi^{01}| = |S_1-S_0|\simeq \delta \omega\simeq 1.5$~MHz as expected for the optimal dispersive readout conditions~\cite{Blais2004}. We emphasize that by reaching the optimal dispersive shift we expect high readout contrast and no additional limitations associated with the relatively small detuning and the system-resonator coupling are imposed on the readout.

As a first experimental test we perform a pulsed spectroscopy of the cavity to identify its resonance frequency for the ground, first and second states of the qutrit. The qutrit is prepared in its energy basis states by applying the control pulses from the set \{$I$, $R^{01}_x(\pi)$, $R^{12}_x(\pi) \cdot R^{01}_x(\pi)$\}. Here $R_{\hat n}^{i,i+1}(\phi)$ is a rotation of angle $\phi$ about the axis $\hat n$ in the qutrit subspace spanned by $\{|i\rangle, |i+1\rangle\}$, $I$ stands for no control pulses and the rightmost pulse in a sequence is applied first in time. The rotations are implemented with calibrated Gaussian-shaped microwave pulses resonant with the corresponding transitions applied to the input port of the cavity. After the qutrit logical states are prepared, we apply a square microwave pulse at a frequency close to the resonance frequency of the cavity for several microseconds. The microwaves transmitted through the cavity are amplified by a chain of amplifiers and  their amplitude and phase are detected by a standard heterodyne detection scheme. The amplitude of the digitized signal is integrated over $1.2$~$\mu$s, a time sufficiently long compared to the inverse width of the cavity line but much shorter than the life time of the qutrit and is shown in Fig.~\ref{fig1}(d,e) as a function of frequency. The data demonstrates that the cavity frequencies for the qutrit in the first and second excited states are clearly distinct away from the sweet spot and identical in the sweet spot.

\subsection{Identifying the sweet spot}
This measurement of the dispersive shifts (Fig.~\ref{fig1}(d,e)) confirms our expectations but does not directly reveal the effect of the measurement on the qutrit and does not provide sufficient accuracy to identify the sweet spot. In order to find the sweet spot precisely, we explore the effect of the readout on the coherence of the $1\leftrightarrow2$ transition for different frequencies of the qutrit. The measurement protocol starts with the preparation of the superposition state $(1/\sqrt2)(|1\rangle + |2\rangle)$ by applying $R^{12}_x(\pi/2) \cdot R^{01}_x(\pi)$ control pulses. Subsequently, the prepared state is tomographically reconstructed after a delay of $2$~$\mu$s using the set of tomography pulses $R^{01}_x(\pi) \otimes \{I,\allowbreak R^{12}_x(\pi),\allowbreak R^{12}_x(\pi/2),\allowbreak R^{12}_{-y}(\pi/2)\}$ preceding the microwave readout pulse through the resonator. An additional readout pulse is inserted right after the preparation pulse with its length swept from $0$ to $1.5$~$\mu$s. The tomography results are summarized in Fig.~\ref{find_sweet_spot}, demonstrating the evolution of the off-diagonal density matrix element $\rho_{12}$ with the length of the inserted readout pulse for different detunings. The first point corresponds to the measurement with no additional readout pulse. The deviation of this point from the unit circle $2|\rho_{12}|=1$ is a consequence of the intrinsic dephasing of the qutrit during delay of $2~\mu$s. The dephasing time depends on the detuning bias point with a tendency to be higher close to the maximal frequency of the qutrit where it is protected from low frequency flux noise in the linear order. It also slightly fluctuates from one measurement to another. 

More interestingly, the spiralling trajectories in Fig.~\ref{find_sweet_spot} manifest the two main effects induced by the measurement on the coherence of the $1\leftrightarrow2$ transition. The tangent motion of the trajectories represent the additional phase shift due to the ac-Stark effect when the resonator is populated by readout photons. The same photons also induce dephasing of the transitions manifested into radial motion toward the centre of the plots. Both effects are strongly dependent on the detuning and disappear towards the sweet spot identified at $f_{01} = 6.901$~GHz. 

Equation (\ref{chi12}) predicts that the sweet spot occurs at $\delta = \alpha$, where we assumed $E_j/E_c \gg 1$ and $\alpha = E_c$. Using the exact solution for the transmon energy levels~\cite{Koch2007}, the values for the frequency $f_{01} = 6.901$~GHz and its anharmonicity $\alpha = -314$~MHz, we can determine the charging energy of the transmon more precisely, $E_c = 281$~MHz. This value is very close to the detuning $|\delta| = 278$~MHz at the sweet spot. Another observation is that the sweet spot slightly shifts in frequency (on the order of a few MHz) for different readout powers. Yet for any power within the dispersive approximation it is possible to find conditions with fully identical responses for states $|1\rangle$ and $|2\rangle$ and a negligible effect of the readout on the coherence of the $1\leftrightarrow2$ transition.

\begin{figure}[htbp]
\begin{center}
\includegraphics[width=0.4\textwidth]{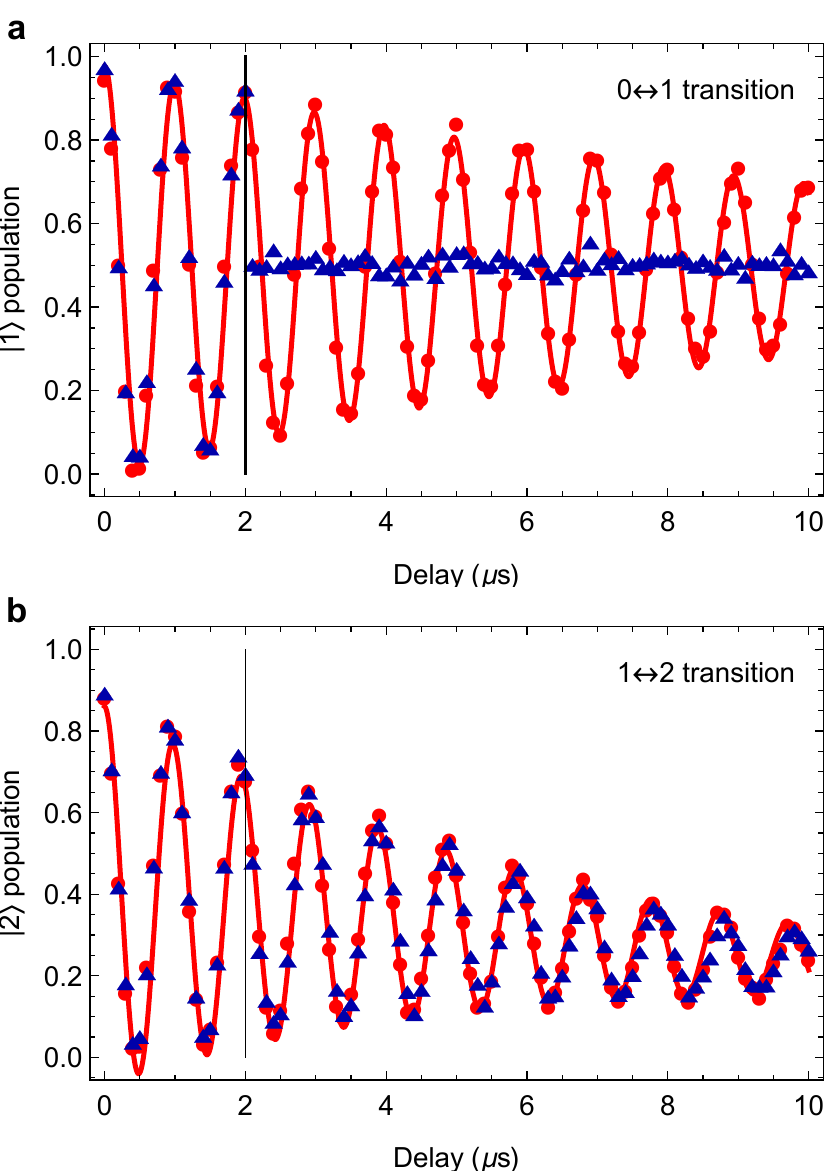}
\caption{
	Effect of the readout on Ramsey fringes of the $0\leftrightarrow1$ and $1\leftrightarrow2$ transitions. 
	(a) Measurement of conventional Ramsey fringes (red dots) for the $0\leftrightarrow1$ transition performed with two $R^{01}_{x}(\pi/2)$ pulses separated by a delay. The solid red line shows a fit to the expected dependence $(1/2)(\exp(-t/T^{01}_2)\cos\delta\omega t+1)$ with $T^{01}_2 = 11.2$~$\mu$s. The dark blue triangles represent a measurement with an additional readout inserted between the pulses for delays exceeding $2~\mu$s. The solid vertical line indicates the delay for which the additional readout first appears. 
	(b) Ramsey fringes for the $1\leftrightarrow2$ transition. Experimental data (red dots) are fitted to  $(1/2)(\exp(-t/T^{12}_2)\cos\delta\omega t+ B\exp(-t/T_1^{01})+ C)$ where $T^{12}_2 = 5.78~\mu$s is the decoherence time for the $1\leftrightarrow2$ transition, the second term accounts for relaxation of the first excited state with $T_1^{01}=15.0~\mu$s and $C = 0.078$ is the thermal occupation of the first excited state. As in a, the dark blue triangles represents a measurement with an additional readout. No effect of the additional readout is visible on the coherence of the $1\leftrightarrow2$ transition.
} 
\label{Ramsey}
\end{center}
\end{figure}

\subsection{Effect of the readout on the qutrit state}
With the sweet spot condition precisely identified, we characterize the effect of the readout in more detail. 
First, for demonstration purposes, we perform Ramsey fringe experiments for the $0\leftrightarrow1$ and $1\leftrightarrow2$ transitions with and without an additional readout (Fig.~\ref{Ramsey}(a,b)). 

To measure Ramsey fringes for the $0\leftrightarrow1$ transition we apply a $R^{01}_{x}(\pi/2)$ pulse to prepare the superposition $(1/\sqrt 2) (|0\rangle+|1\rangle)$. After a delay, another $R^{01}_{x}(\pi/2)$ pulse is applied, followed by the readout pulse. Both pulses were detuned by $\sim 1$~MHz from the $0\leftrightarrow1$ transition to reveal precession of the prepared state relative to the rotating frame defined by the driving frequency.
The red curve in Fig.~\ref{Ramsey}(a) shows standard Ramsey oscillations with a decay time of $T_2^{01} = 11.2~\mu$s. The dark blue curve shows Ramsey oscillations where an additional readout pulse is inserted between the $\pi/2$ pulses for delays exceeding $2$~$\mu$s. 
The dark blue curve shows Ramsey oscillations where an additional readout pulse is inserted for delays exceeding 2 us. It starts directly after the first $\pi/2$
 pulse and ends 2~$\mu$s before the second one to allow ample time for the cavity to ring down before the second $\pi/2$ pulse. An additional readout pulse of 150~ns duration is sufficient to completely erase the $0\leftrightarrow1$ coherence. 

For the $1\leftrightarrow2$ transition we first prepare the superposition state $(1/\sqrt2)(|1\rangle + |2\rangle)$ with $R^{12}_x(\pi/2) \cdot R^{01}_x(\pi)$ pulses. After a delay, the coherence of $\rho_{12}$ is converted into a population of the second excited state $|2\rangle$ by another control pulse $R_{x}^{12}(\pi/2)$. The populations of the qutrit are then reconstructed using a set of tomography pulses \{$I$, $R^{01}_x(\pi)$, $R^{12}_x(\pi) \cdot R^{01}_x(\pi)$\} placed in front of the readout pulse. The population of $|2\rangle$ exhibits an expected decay curve with $T^{12}_2 = 5.77$~$\mu$s and with additional terms accounting for relaxation of the first excited state and its thermal occupation. As already anticipated from Fig.~\ref{find_sweet_spot}, there is no visible effect of the additional readout on the coherence of this transition.

As an even more representative example we prepare the state $|\psi_0\rangle = (1/\sqrt 3)(|0\rangle +|1\rangle +|2\rangle)$ and perform its full tomography after a delay of $450$~ns. To accumulate enough knowledge to reconstruct a density operator nine measurements are performed, each with a particular combination of control microwave pulses in front of the measurement pulse through the readout cavity. It is interesting to note that being insensitive to the relative populations of the states $|1\rangle$ and $|2\rangle$ our readout realizes a specific pathological case where the previously used scheme for qutrit tomography~\cite{Bianchetti2010} fails as it is not tomographically complete. To correct for this pathology we use a modified set of the tomographic pulses shown in Table~\ref{tab: tomo rotations}. Physical density matrices are reconstructed from the measured responses through a maximum likelihood reconstruction by semidefinite programming~\cite{Vandenberghe1996}.

\begin{table}[tb]
		\begin{tabular}{c|ccccc|c}
			i &  & & $U_i$ && & $|\psi_i\rangle$ \\
			\hline
			1 &  &  &$I$ & & & $|0\rangle$  \\
			2 & & &$R^{01}_x(\pi/2)$& & & $(|0\rangle + |1\rangle)/\sqrt2$  \\
			3 & & &$R^{01}_y(\pi/2)$& & & $(|0\rangle - i|1\rangle)/\sqrt2$ \\
			4 & & &$R^{01}_x(\pi)$  & & & $|1\rangle$ \\
			5 & & & $R^{12}_x(\pi/2)$  & $\cdot$ & $R^{01}_x(\pi)$ & $(|1\rangle + |2\rangle)/\sqrt2$\\		
			6 & & & $R^{12}_y(\pi/2)$  & $\cdot$ & $R^{01}_x(\pi)$ & $(|1\rangle - i|2\rangle)/\sqrt2$\\
 			7 & $R^{01}_x(\pi)$    & $\cdot$ & $R^{12}_x(\pi/2)$ &$\cdot$ & $R^{01}_x(\pi)$ & $|2\rangle$ \\
			8 & $R^{01}_x(\pi)$    & $\cdot$ & $R^{12}_y(\pi/2)$ &$\cdot$ & $R^{01}_x(\pi)$ & $(|0\rangle + |2\rangle)/\sqrt2$ \\
			9 & $R^{01}_x(\pi)$    & $\cdot$ & $R^{12}_x(\pi)$   &$\cdot$ & $R^{01}_x(\pi)$ & $(|0\rangle + i|2\rangle)/\sqrt2$\\	
		\end{tabular}
	\caption{The set of tomography pulses $U_i$ for a qutrit sufficient to reconstruct an arbitrary state density operator when the first and second excited states are indistinguishable. The set of initial states $|\psi_i\rangle$ which can be prepared and reconstructed for the full process tomography.}
	\label{tab: tomo rotations}
\end{table}

\begin{figure}[htbp]
\begin{center}
\includegraphics[width=\columnwidth, trim=0in 0.68in 0in 0.35in]{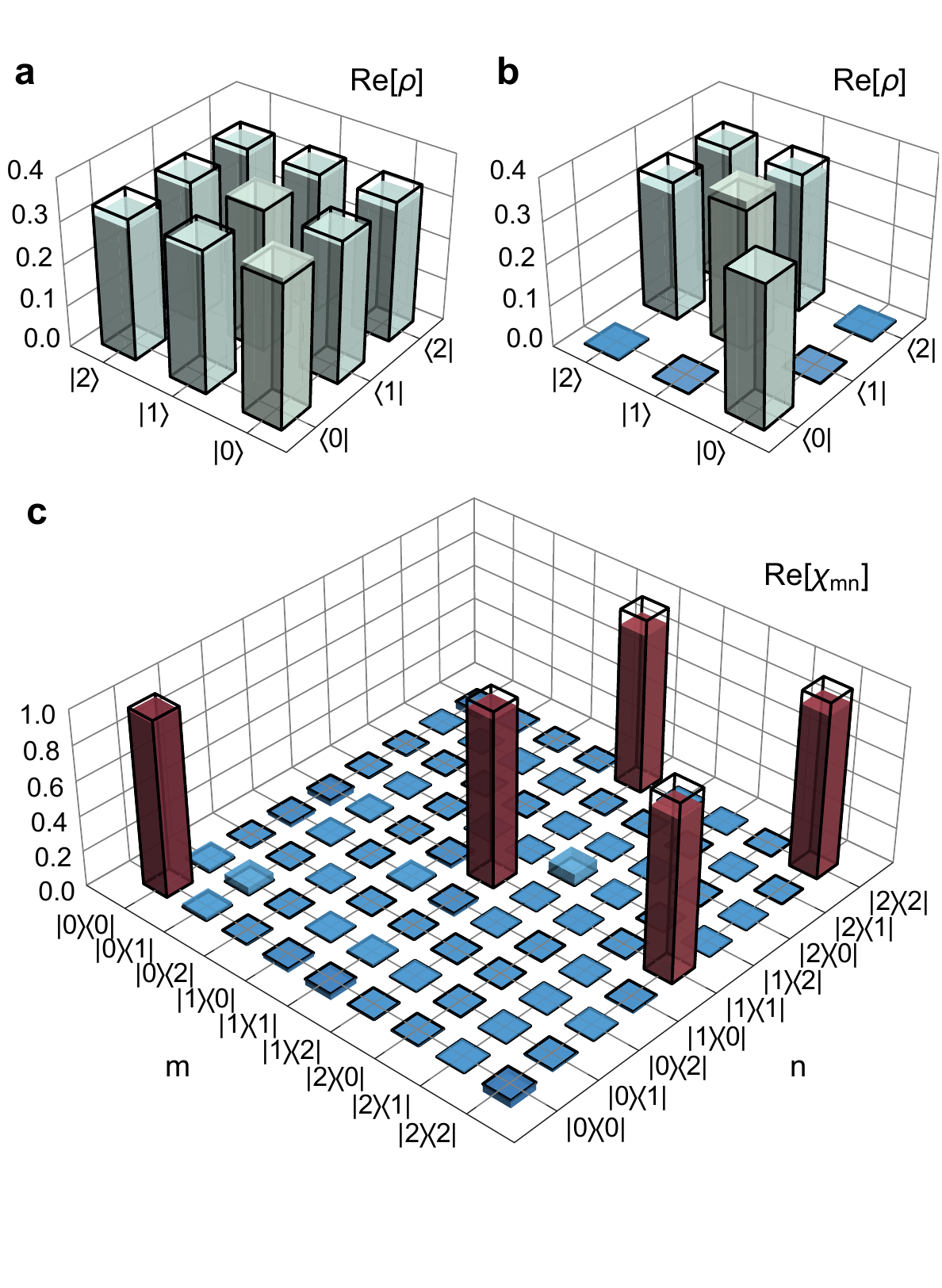}
\caption{
	(a) The reconstructed density operator $\rho_{\rm without}$ after a delay of $450$~ns. 
	(b) The reconstructed density operator $\rho_{\rm with}$ a after delay of $450$~ns with an additional readout pulse of $150$~ns between the preparation and tomography pulses. Wire frames indicate the density operators $\rho_{\rm b} = |\psi_0\rangle\langle \psi_0|$ and $\rho_{\rm a} = \sum_{i=\pm} M_i \rho_{\rm b} M_i^{\dagger}$, respectively. 
	(c) The process matrix $\chi_{\rm meas}$ for the readout. The fidelity of the measured process to the process matrix for ideal projective measurement described by  $\{M_{+1},M_{-1}\}=\{|0\rangle\langle 0|,I - |0\rangle\langle 0|\}$ (indicated by the wire frame) is $94.3\%$. The reduction of the fidelity is attributed to intrinsic decoherence of the qutrit during the time of the readout.} 
\label{fig4}
\end{center}
\end{figure}

To highlight the effect of the readout we perform state tomography without (Fig.~\ref{fig4}(a)) and with (Fig.~\ref{fig4}(b)) an additional readout pulse of 150~ns between the preparation and tomography pulses. An additional delay time of 300~ns was needed to let the cavity ring down to avoid interference with the tomography pulses. As expected, we observe the total decay of the coherences $\rho_{01}$ and $\rho_{02}$ as a result of the collapse of the wave-function due to projection onto state $|0\rangle$, while the coherence in the orthogonal subspace spanned by $\{|1\rangle, |2\rangle\}$ remains virtually unchanged (see Fig. 4b). In addition, the populations of the qutrit also remain virtually unchanged which shows the quantum non-demolition character of the readout. We obtain fidelities of $F = {\rm Tr} \left[ \sqrt{ \sqrt{\rho_{\rm b}} \rho_{\rm without} \sqrt{\rho_{\rm b}} } \right]^2 = 97.1\%$ without the additional readout and $F = {\rm Tr} \left[ \sqrt{ \sqrt{\rho_{\rm a}} \rho_{\rm with} \sqrt{\rho_{\rm a}} } \right]^2 =  96.9\%$ with the additional readout, where $\rho_{\rm b} = |\psi_0\rangle\langle \psi_0|$ and $\rho_{\rm a} = \sum_{i=\pm} M_i \rho_{\rm b} M_i^{\dagger}$ with $\{M_{+},M_{-}\}=\{|0\rangle\langle 0|,I - |0\rangle\langle 0|\}$. The slight change in the populations and decay of $\rho_{12}$ can be fully attributed to the effect of the intrinsic relaxation and dephasing of the qutrit during the readout and delay times.

For a quantitative analysis we perform process tomography~\cite{Nielsen2000} and reconstruct the process matrix, $\chi_{\rm meas}$ of the readout. For that purpose, we prepare a set of states shown in Table~\ref{tab: tomo rotations} and perform state tomography on the respective output states after the readout pulse. A maximum likelihood method similar to the one used to reconstruct density matrices is employed to obtain the physical process matrix that best describes the measured data. The process matrix (Fig.~\ref{fig4}(c)) shows the key features of the expected ideal projective measurement generated by the set of measurement operators $\{M_{\pm}\}$. We find a process fidelity~\cite{Gilchrist2005} of $F = (1/9) \left[ {\rm Tr} \sqrt{ \sqrt{\chi_{\rm ideal}} \chi_{\rm meas} \sqrt{\chi_{\rm ideal}} } \right]^2 = 94.3\%$.

\subsection{Measurement compatibility and application to testing of quantum contextuality}
Once an experimentalist identifies a physical mechanism to perform a projection on state $|0\rangle$ described by $\{|0\rangle\langle 0|,I-|0\rangle \langle 0|\}$ (or on some other state) it is straightforward to realize a projection on an arbitrary state $|v\rangle$ by performing additional manipulations of the state before and after the measurement. More explicitly, if $|v\rangle = U|0\rangle$ it suffices to perform a rotation $U^{\dagger}$  before the measurement  and restore the coordinate system by applying an additional rotation $U$ after the measurement. This measurement procedure is fully equivalent to the measurement of $\rho_{\rm b}$ described by $\{|v\rangle\langle v|,I-|v\rangle \langle v|\}$, as required.

If two states are orthogonal, $\langle u |v\rangle = 0$, their corresponding observables commute, $[A_v, A_u] = 0$, and are expected to be compatible: for any combination (sequence) of these two measurements the values of $A_v$ and $A_u$ agree independently of their position in the sequence~\cite{Szangolies2015,Guhne2010}. One of the practical ways to test compatibility is to establish a bound~\cite{Guhne2010}
\begin{equation}
|\langle A_u|A_u A_v\rangle - \langle A_u|A_v A_u\rangle| \leq \varepsilon_{uv},
\end{equation}
for all possible $\rho_{\rm b}$. Here, $\langle A_u|A_u A_v\rangle$ ($\langle A_u|A_v A_u\rangle$) are the expectation values of $A_u$ for two sequential measurements when $A_v$ is measured after (before) $A_u$. Using $\langle u |v\rangle = 0$ it is easy to show that indeed $\varepsilon_{uv} = 0$ if measurements are described by $\{|i\rangle\langle i|, I - |i\rangle\langle i|\}_{i=u,v}$. 

Let us consider the measurement described by the set of operators $\{|0\rangle\langle 0|, |1\rangle\langle 1|, |2\rangle\langle 2|\}$ and corresponding outcomes $\{-1, 1, 1\}$. One can transform this measurement to $\{ |\psi_1\rangle\langle \psi_1|, |0\rangle\langle 0|, |\psi_2\rangle\langle \psi_2|\}$ where $|\psi_{1,2}\rangle = (1/\sqrt2) (|1\rangle \pm |2\rangle)$ 
with two control rotations before the measurement $R^{01}_{-y}(\pi) \cdot R^{12}_{-y}(\pi/2)$ and with the inverse rotations after the measurement. Being measured independently both procedures will reproduce the outcomes of the corresponding binary-outcome compatible measurements $ A_0$ and $A_{\psi_1}$. However, despite orthogonality, $\langle 0|\psi_1\rangle = 0$, some of the other measurement operators do not commute: $\langle 1|\psi_{1,2}\rangle \neq 0$, $\langle 2|\psi_{1,2}\rangle \neq 0$ and within a sequence the outcomes of the measurement will not agree. As the most profound example one can consider $\rho_{\rm b} = |\psi_1\rangle\langle\psi_1|$ which yields $\langle A_{\psi_1}\rangle=\langle A_{\psi_1}|A_{\psi_1} A_0\rangle = 1$, $\langle A_{\psi_1}|A_0 A_{\psi_1} \rangle = 0$ showing a strong disturbance of one measurement by the other. Using the reconstructed $\chi$-matrix and assuming perfect control pulses and perfect readout contrast we can evaluate the degree of incompatibility of our measurement as $\varepsilon_{0\psi_1} = 0.08$.

The Kochen-Specker contextuality test scenario~\cite{Kochen1967} involves a number of observables $\{A_{v_i}\}$ and a set of indices $\wp$, such that for $i$, $j \in \wp$: $\langle v_i|v_j\rangle = 0$.  Based on the assumption of compatibility of the corresponding $A_{v_i}$ and $A_{v_j}$, one can derive experimentally testable inequalities which in the simplest cases require measurement of two-observable correlations $\langle A_{v_i} A_{v_j}\rangle$~(see, for example, \cite{Klyachko2008,Yu2012}). 
For imperfectly compatible measurements the Kochen-Specker inequalities can be extended through introducing additional ‘error’ terms $\varepsilon_{v_iv_j}$
to compensate for possible imperfections~\cite{Szangolies2015,Guhne2010}. Further increase of compatibility can be achieved by using qutrits with longer coherence times~\cite{Riste2012,Rigetti2012}.



\section{Summary and Discussions}
In conclusion, we used a multi-level superconducting quantum system coupled to a microwave cavity to engineer a regime where the dispersive shifts of the cavity are identical for the first and second excited states of the qutrit. This regime allowed us to realize a binary-outcome projective measurement of a superconducting qutrit on its ground state with a process fidelity of $94.3\%$. The distinctive property of this measurement is the partial projection of the qutrit space onto its ground state with preservation of quantum coherence between its excited states.
Complemented with standard microwave control, this measurement scheme can be used to reveal and study the contextual nature of superconducting circuits. 

Another important application for our readout scheme is leakage error detection relevant to error correction within transmon-resonator based architectures~\cite{Ghosh2012}. Transmon qubits are known to be particularly susceptible to leakage errors due to their weakly anharmonic level structure~\cite{Motzoi2009}. Unlike errors induced by decoherence, leakage errors are not directly handled by conventional error correction schemes and require additional care~\cite{Suchara2015}. Our measurement scheme enables the direct quantum non-demolition measurement of leakage errors similar to the quantum non-demolition measurement of qubit loss  for neutral atoms in an addressable optical lattice~\cite{Vala2005}. 

The experiment is also an instructive demonstration of the concept of a decoherence free subspace for an indivisible multilevel system. Due to the very same mechanism which protects the qutrit subspace~\cite{Lidar2003} during the measurement, a logical qubit encoded in the first and second excited states of the physical qutrit will be protected from the dephasing induced by photons in the cavity, one of the major sources of dephasing for superconducting qubits~\cite{Sears2012}. 

The readout scheme presented here uses only the most basic elements of circuit QED and can easily be incorporated into other circuit QED systems. It relies only on the anharmonicity of the multilevel system coupled to a harmonic oscillator and can potentially be applied to other physical implementations.

\begin{acknowledgements}
We thank Clemens M\"uller, Fabio Costa, Nathan Langford, Stephen Bartlett and Tim Ralph for useful discussions and Kirill Shulga for his help with the measurements. MJ, ARH, PM were supported by the Australian Research Council Centre of Excellence CE110001013. YR was supported by the Discovery Project DP150101033. AF was supported by the ARC Future Fellowship FT140100338. KJ was supported by the CCQED network. The superconducting qutrit was fabricated at CEA Saclay, France by KJ. We would like to give special thanks to Denis Vion for providing the qutrit sample for the experiment.
\end{acknowledgements}

\bibliographystyle{apsrev4-1}
\bibliography{Z:/RefDB/SQDRefDB}

\end{document}